# Journal Name

## ARTICLE TYPE

# Interactive Human - Machine Learning Framework for Modelling of Ferroelectric-Dielectric Composites†

Ning Liu,‡[a] Achintha Ihalage,¶[a] Hangfeng Zhang,[b] Henry Giddens,[a] Haixue Yan,[b] and Yang Hao*[a]

Data driven materials discovery and optimization requires databases that are error free and experimentally verified. Performing material measurements are time-consuming and often restricted by the fact that material sample preparations are non-trivial, labour-intensive and expensive. Numerical modelling of materials has been studied over the years in order to address these issues and nowadays it has been developed at multi-scale and multi-physics levels. However, numerical models for nano-composites, especially for ferroelectrics are limited due to multiple unknowns including oxygen vacancy densities, grain sizes and domain boundaries existing in the system. In this work, we introduce a human-machine interactive learning framework by developing a scalable semi-empirical model to accurately predict material properties enabled by deep learning (DL). MgO-doped BST ($Ba_xSr_{1-x}TiO_3$) is selected as an example ferroelectric-dielectric composite for validation. The DL model transfer-learns the experimental features of materials from a measurement database which includes data for over 100 different ferroelectric composites collected by screening the published data and combining our own measurement data. The trained DL model is utilized in providing feedback to human researchers, who then refine computer model parameters accordingly, hence completing the interactive learning cycle. Finally, the developed DL model is applied to predict and optimise new ferroelectric-dielectric composites with the highest figure of merit (FOM) value.

*[a] School of Electronic Engineering and Computer Science, Queen Mary University of London, E1 4NS, United Kingdom. E-mail: y.hao@qmul.ac.uk*
*[b] School of Engineering and Materials Science, Queen Mary University of London, E1 4NS, United Kingdom.*
† Electronic Supplementary Information (ESI) available: Full model derivation and analysis are available as supplementary text. Simulated and measurement materials databases are also provided. See DOI: 00.0000/00000000.
‡¶ These authors contributed equally to this work.

## Introduction

Materials modelling is a key pre-design strategy used to eliminate trial-and-error loops in new materials development process. A range of modelling approaches have been proposed for atomistic and theoretical modelling of materials, such as density functional theory (DFT),[1,2] molecular dynamics (MD),[3,4] Monte Carlo method,[5,6] semi-empirical physical models[7,8] and finite element models.[9,10] Modelling of spontaneous polarization, dielectric properties, ferroelectric to paraelectric phase transition taking effect at Curie point ($T_c$) and structural properties of ferroelectric materials have been extensively scrutinized with both atomic level and numerical simulations.[11–13] With the advancement of material characterization techniques in later years, some of the specific behaviours such as domain-wall motions, defects causing the pinning effect, negative capacitance and the presence of local dipole components in the paraelectric region have been experimentally identified and been considered in the theoretical models.[14–18] However, the first principle calculations are computationally expensive and theoretical modelling requires a profound knowledge on the physical phenomena of the material. Hence, machine learning (ML) is increasingly being used to effectively bypass these calculations.[19–21] Thus, these ML frameworks are used to predict material properties[22,23] and to design and discover novel materials.[24,25] The integration of ML with theoretical models, despite being useful for materials discovery and optimization, still remains as a less-explored research field.

The real challenge is therefore to develop models that comply well with measurement data.[26] In this work, we propose a human-machine interactive learning framework that a new computer model based on semi-empirical calculations of ferroelectrics is developed to model ferroelectric-dielectric composites by feeding 'machine-learned' experimental features in order to significantly boost the modelling accuracy. The proposed framework is embraced by the power of inherent natures of learning abilities where ML algorithms are good at rapid learning from mass data while capturing even the slightest variation, and humans are empowered by their analytical knowledge to anticipate new scenarios by abstracting different domains.

To demonstrate the concept as shown in Fig. 1, we select a semi-empirical model for ferroelectrics, known as Vendik model developed from Landau-Ginzburg theory[7,8,27–29] and $Ba_xSr_{1−x}TiO_3$ (BST) as a modelling example. The Vendik model allows us to calculate dielectric properties of both incipient and displacive types of ferroelectrics as a function of temperature, biasing field, frequency and material defects. However, by comparing our initial calculations with our measurement data and those from[30] on BST ceramics, we have concluded that the model overestimates dielectric constants at high frequencies (over 1GHz). Furthermore, the original Vendik model is not designed for ferroelectric-dielectric composites, which are commonly synthesised by research scientists to modify material parameters such as dielectric constant, loss tangent and tunability.

In the present study, we first theoretically develop an improved Vendik model which is valid for high frequencies as well as ferroelectric-dielectric composites such as MgO-doped BST ceramics by considering the new mechanisms brought by the dielectric dopant and further reflecting them on analytical equations. Deep learning (DL), which is a branch of ML will then be used to successively learn from a simulated database and an experimental database. We will receive feedback from the trained DL model and will subsequently make appropriate refinements to the semi-empirical model parameters such that resulting simulation data adhere to those from measurements. A fully connected deep neural network (DNN) architecture is proposed to avoid the phenomenon known as *catastrophic forgetting*,[31] frequently occurring in the context of *transfer learning*. The tendency of the neural networks to forget what it had learned previously upon learning new information is known as catastrophic forgetting. Learning from two databases can be mapped into a transfer learning task, where the same ML model trained on the simulated database is re-purposed by the subsequent training with the experimental database. The refined theoretical model following the above interactive learning process is experimentally validated with different $Ba_{0.6}Sr_{0.4}TiO_3$ (BST64) samples and the trained DL model is utilized for optimized material modelling to explore the required conditions for designing highly tunable, low loss ferroelectrics operating in the paraelectric state. Combined ML and theoretical modelling supported by the experiments demonstrates the applicability and scalability of the proposed interactive learning framework, seemingly having a vast scope of applications in materials modelling.

## Theory and design

**Original Vendik model**

An analytic equation to calculate the complex dielectric constant of ferroelectric materials under different temperatures and electric fields for both ferroelectric and paraelectric states was proposed by Vendik et al.[7,8] The equations are derived based on the conventional Landau theory and four energy dissipation mechanisms are considered in the derivation (see Appendix, section A.1 for full model derivation). Thus, the proposed equation to calculate the complex permittivity of $Ba_xSr_{1−x}TiO_3$ ceramics can be formulated as: $\varepsilon(E,T,f,x,\xi_s) = \frac{\varepsilon_{00}(x)}{G(E,T,x,\xi_s)^{-1}+\sum_{q=1}^{4}\Gamma_q(E,T,f,x,\xi_s)}$ (1), where $G(E,T,x,\xi_s)$ is the real part of the Green function for a dielectric response of the ferroelectric, $x$ is the barium proportion, $T$ is the temperature and $f$ is the operating frequency of biasing field $E$. $\varepsilon_{00}(x)$ is an analogue of Curie-Weiss constant $C$ and can be represented as $\varepsilon_{00} = C/T_c$. $\xi_s$ is the statistical dispersion of the biasing field (also known as defect factor) which reflects the 'quality' of the material and corresponds to defects (including oxygen vacancies and inhomogeneity) in the material. $\Gamma_{1,2,3,4}$ refer to the four energy dissipation (loss) mechanisms considered in the original model (see Appendix, section A.1 for the detailed discussion on $\xi_s$ and $\Gamma_{1,2,3,4}$). Therefore, the dielectric loss, i.e loss tangent of a ferroelectric material could be expressed as: $\tan\delta = \frac{Im[\varepsilon(E,T,f,x,\xi_s)]}{Re[\varepsilon(E,T,f,x,\xi_s)]}$ (2). All model constants in the current model are set to be identical with those of the original Vendik model[8] and the constants referring to BST material are tabulated in Appendix, section A.3.

**Improved Vendik model for high frequencies**

By comparing simulations from the original model with our measurement data for BST at high radio frequencies (>1GHz), it was observed that the original Vendik model needs to be refined to apply for high frequencies. For example, as can be seen from Fig. 2, between 8 GHz and 12 GHz, BST64 displays a permittivity around 50 depending on different synthesis conditions whereas the original model simulations (with defect parameter $\xi_s = 0.8$) yield the values in order of thousands. This inaccuracy can be attributed to the fact that the contribution of the reduced polarization at high frequencies in ferroelectrics was not considered in the original model.

For a typical dielectric placed in an electric field $E$ between two flat electrodes, the relationship between the internal polarization $P_{int}$ and dielectric constant $\varepsilon_r$ is described as: $\varepsilon_r = 1 + \frac{P_{int}}{\varepsilon_0 E}$ (3), where $\varepsilon_0$ represents the permittivity of free space. This basic equation tells, at such condition, $\varepsilon_r$ is positively proportional to internal polarization inside the dielectric. Therefore, at high frequencies, the permittivity of BST will be much reduced as dipoles are unable to respond to the changing directions of alternating field whereas at intermediate frequencies, the dipoles can partially reorient with the changing of alternating field direction, but will increasingly lag behind as the frequency increases.[32] Specifically at the ferroelectric phase, polarization in BST material will be reduced with increasing frequency of biasing voltage as domain wall motion cannot follow the alternating field[33] and/or at cryogenic temperatures where domain wall motion is thermally frozen.[33]

Having studied measurement data published in[34,35] and our own data on BST64, we expect, for pure BST material, a steady high permittivity at low frequencies (1kHz-1MHz) and a dramatic drop near 1GHz. By considering the firm dependence of the permittivity on frequency, we introduce a modified Vendik model with: $\varepsilon(E,T,f,x,\xi_s) = \frac{\varepsilon_{00}(x)}{[G(E,T,x,\xi_s)K(f)]^{-1}+\sum_{q=1}^{4}\Gamma_q(E,T,f,x,\xi_s)}$ (4), where $K(f)$ is written as: $K(f) = k_a \tanh[k_b \ln(f) + k_c] + k_d$ (5). The constants $k_a, k_b, k_c, k_d$ in equation (5) relating to the modified model were found by a curve fitting process between our measurement data on BST64. The model constants were found to be -0.442, 0.490, -3.2 and 0.453 respectively and the details of curve fitting can be referred in Appendix, section B.2 Fig. B.7. After the introduction of frequency-dependent factor $K(f)$, calculation results are now comparable with those from measurements, as can be observed in Fig. 2.

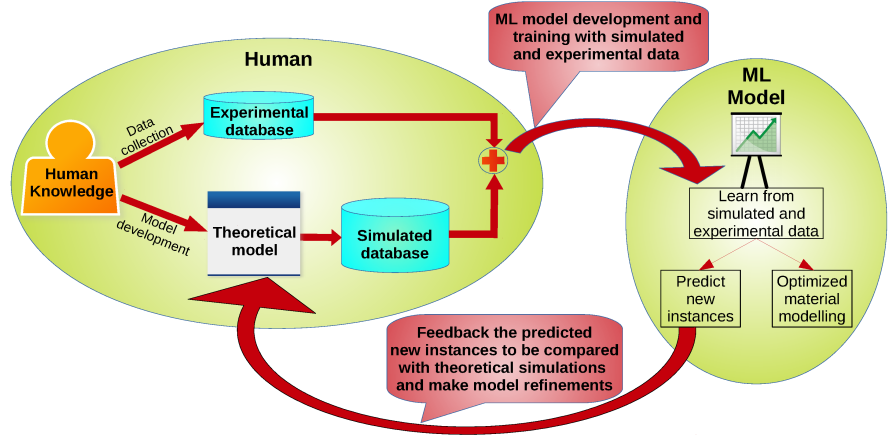

**Fig. 1** Human-machine interactive learning framework. Human develops a theoretical model based on the knowledge. The developed model is used to produce a big database which enables "materials by design". An experimental dataset is also created by screening those from published literature and combining those with our own measurements. A machine learning model successively learns from these two databases and predicts new instances. Predicted results are compared with theoretical calculations and further refinements are then introduced to the theoretical model such that final numerical results reflect the actual behaviour of materials.

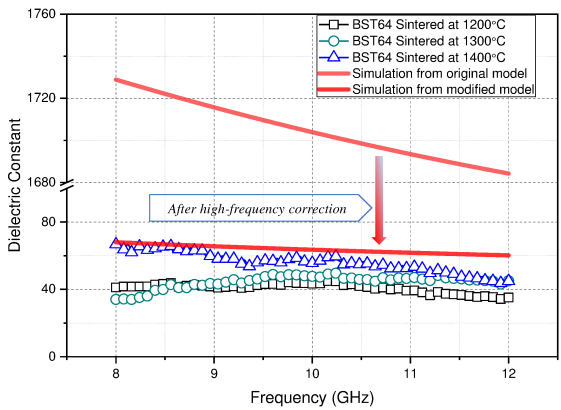

**Fig. 2** Experimentally obtained permittivity of BST64 ceramics between 8 to 12 GHz synthesized at different sintering temperatures, measured at room temperature, comparing to simulation data from both original and modified models, where $\xi_s$ is set to 0.8.

**Ferroelectric-dielectric composite modelling**

MgO doping is one of the most prevalent methods being used to alter the dielectric properties of BST ceramics, especially because the formed composition still maintains the $ABO_3$ perovskite structure and the macroscopic ferroelectric behaviour at low concentration of MgO doping. For instance, a single-phase solid solution can be achieved at MgO doping levels up to 5 mol% whereas with increasing MgO concentration, multi-phase composites are obtained at 20 mol% MgO, confirming the emergence of BST-MgO interfaces in the material.[36] This will give rise to charged defects at the interfaces corresponding to higher extrinsic losses in the composite especially at low frequencies.[37] Based on the above improved model and taking MgO-doped BST as an example, we develop a semi-empirical model for ferroelectric-dielectric 'composites', for the sake of convenience, which includes the single-phase BST-MgO composition with low doping levels of MgO as well.

With MgO doping, initially the smaller B-site $Ti^{4+}$ ions get replaced by larger $Mg^{2+}$ ions and the oxygen vacancies will increase causing more defects resulting in a reduced dielectric constant of the material. Further increment in MgO content results in A-site $Ba^{2+}$ ions being replaced by $Mg^{2+}$ due to the high oxygen vacancies and this further reduces the dielectric constant.[38] Moreover, even at paraelectric phase, polar regions are still identified in BST material and the movement of these regions will also be restricted by new complexes brought by MgO doping which in turns decreases the dielectric constant in the paraelectric phase. As mentioned, in multi-phase composites, the BST-MgO interfaces will introduce more charged defects in the material. As a consequence, the value of $\xi_s$ related to the defects in the material needs to be elevated. Therefore, another term $\xi_{Mg}$, which is positively correlated with MgO doping content is added to the previous parameter $\xi_s$, where the new defect parameter $\xi'_s$ could be presented as: $\xi'_s = \xi_s + \xi_{Mg}$ (6).

The replacement of $Ti^{4+}$ ions by $Mg^{2+}$ suppresses the domain-wall motion which in turns reduce the dielectric loss because the loss originated by domain-wall motions is quite significant, specially near the $T_c$. In contrast, the defects caused by oxygen vacancies result in a slight increase of the loss at high temperatures, generally above 350K for BST-MgO composites.[38] Therefore, we introduce a factor $K_{Mg}$ which is dependent on MgO content $\xi_{Mg}$ as: $K_{Mg} = \exp(-c\xi_{Mg})$ (7). In the developed ferroelectric-dielectric composite model, the equations relating to the loss tan-gent of the material are scaled by a factor of $K_{Mg}$ and the previous defect parameter ($\xi_s$) is substituted with the new parameter $\xi'_s$. The value of $c$ is positive and determines the dependence of $K_{Mg}$ on doping content factor $\xi_{Mg}$ and for the sake of convenience, we assume $c = 1$ for the present model. More detailed explanation on analytic equations reflecting the changes brought by MgO doping can be referred in Appendix, section B.2.

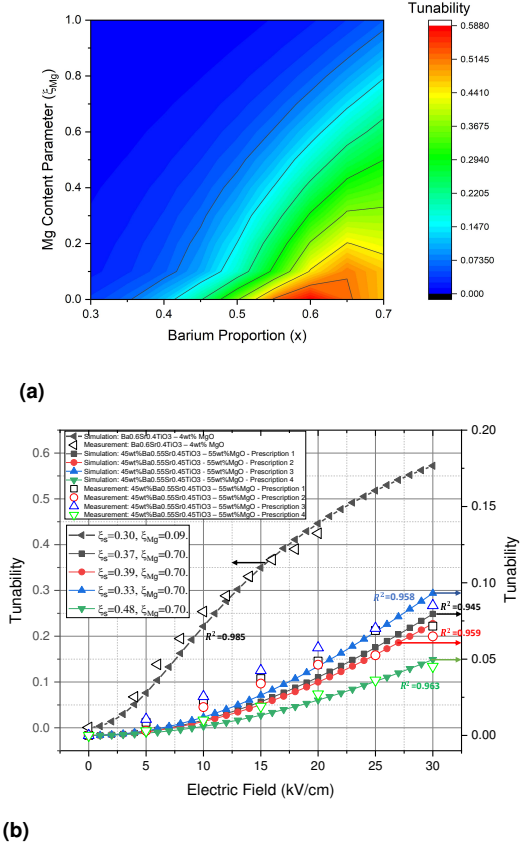

**Fig. 3** (a) Contour plot of room temperature tunability at 20 kV/cm calculated by the modified model, plotting versus barium proportions from 0.3 to 0.7, MgO content parameter $\xi_{Mg}$, from 0 to 1; (b) Simulated tunability curves fitting with measurement data of different BST-MgO composites (extracted from previous literatures:[37,39] $Ba_{0.6}Sr_{0.4}TiO_3$-4wt% MgO and 4 different prescriptions of 45wt%$Ba_{0.55}Sr_{0.45}TiO_3$-55wt%MgO), obtained with the highest R-square values respectively. The direction of the arrows pointing to indicates the referred y-axis for each set of data. For each set of simulated curves, defect parameter $\xi_s$ and Mg content parameter $\xi_{Mg}$ were calculated respectively.

The contour plot in Fig. 3a presents the simulation results of the tunability of BST-MgO composites with various barium proportions and MgO contents as obtained by the developed model. In the plot, with increasing MgO content, we can observe a lower tunability from the composite. Moreover, we fit the measured tunability data of BST-MgO materials (lightly-doped BST64 and heavily-doped composites of 4 different prescriptions) extracted from[37,39] with our modified theoretical model simulations. For the curve fitting process, the best fitting was obtained with optimum values of defect parameter $\xi_s$ and Mg content parameter $\xi_{Mg}$, as shown in Fig. 3b. For lightly-doped BST64 ($Ba_{0.6}Sr_{0.4}TiO_3$-4wt% MgO), we have a very low MgO parameter calculated at 0.09. When heavily doped, the BST-MgO composites (45wt%$Ba_{0.55}Sr_{0.45}TiO_3$-55wt%MgO) become much less tunable, smaller than 10% for all four different prescriptions. We assume MgO content parameters does not change ($\xi_{Mg} = 0.7$) as the MgO doping level is same in all prescriptions. Hence, different defect parameters were obtained for each prescription of BST composites respectively, which agrees to the theoretical definition for $\xi_s$ since lattice parameters vary between different prescription approaches (see Appendix, section A.2).

### Data collection

We created an experimental database of bulk BST composites by screening the published data. This database contains information such as Curie temperature, grain size, dielectric constant at both $T_c$ and room temperature, tunability and loss tangent values at a given biasing field. The data is spanned from 1kHz to microwave frequencies. We used WebPlotDigitizer[40] to extract the data from the plots wherever the data is presented in graphical format rather than in numerical format and this data is provided separately. Altogether, this measurement database contains over 1000 data points for over 100 different BST composite materials. As focussed in this work, majority of data in this database represented BST-MgO composites, however, other compositions such as BST-$MgAl_2O_4$, BST-$Mg_2TiO_4$ and BST-$MgZrO_3$ were also present. Another database was created from theoretical model simulations. The database comprises tunability and loss tangent values for different barium proportions (0.3≤x≤0.9) of pure BST and MgO-doped BST composites at different $\xi_s$ (0.2≤$\xi_s$≤0.8), $\xi_{Mg}$ (0≤$\xi_{Mg}$≤0.8), electric field (0≤E≤30 kV/cm) and frequency levels (f $\in \{10^i |$ i$\in \mathbb{Z}$:i$\in [3,10]\}$Hz). This simulated database contains around 35000 data points for 35 different BST composites.

## Results and discussion

### Deep learning model

The underlying requirement for developing a theoretical model for materials discovery is to improve its accuracy by incorporating experimental data. Machine learning stands out as the obvious choice of learning from data and here we propose a fully connected DNN that acts as the interface between the theoretical model and the measurement dataset. As shown in Fig. 4, a deep learning model is firstly developed to learn from the database generated from theoretical calculations. It then learns from the measurement database to reflect the actual behaviour of the BST material. The trained DL model is used to predict new ferroelectric-dielectric composites and the predictions are fed back to the human to make appropriate adjustments to the theoretical model parameters. This section is divided into two parts. In the first phase, we propose a suitable DL architecture that can fully emulate the theoretical model and verify the trained DL model predictions with theoretical simulations. In the second phase, we employ the ML concept of transfer learning to retrain the verified DL model with the small measurement dataset in such a way that it preserves what it had learned from the theoretical model, while learning the experimental features.

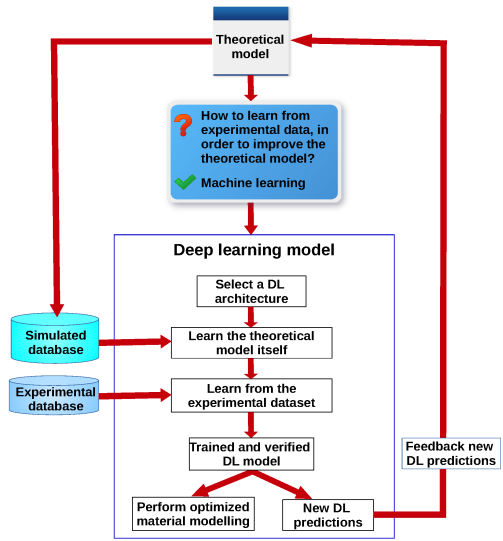

**Fig. 4** Deep learning work flow. DL model first learns the theoretical model itself, followed by the experimental dataset. The accuracy of the trained DL model is quantified and thus it is used to predict new instances that are fed back to the theoretical model where a human could make appropriate adjustments. The DL model is finally used for optimized material modelling.

## Emulation of the theoretical model and verification

Deep learning architecture is one of the determining factors of the final model performance. It is vital to foresee the ultimate objective of using deep learning in the problem before proposing an architecture. It has become clear that higher tunabilities and lower loss tangents are two of the most preferred characteristics in tunable devices. However, by observing the simulations, we noticed that loss tangent estimation from the theoretical model could still be improved whereas the tunability estimation matches well with the measurement data. Hence, we envisage the idea of loss tangent improvement and build the DL model upon that.

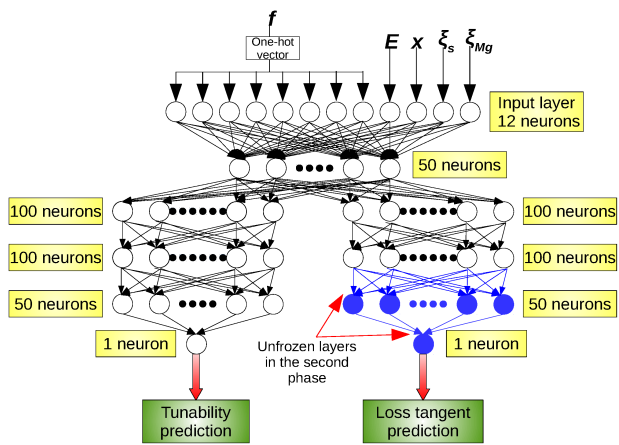

**Fig. 5** Deep neural network architecture. Full model is trained on the simulated database in the first phase. Only the blue-shadowed two layers are trained on the measurement database in the second phase to avoid catastrophic forgetting.

**Table 1** DNN training performance

|  | Training MSE | Validation MSE | $R^2$ Value |
|---|---|---|---|
| Tunability | $2.5 \times 10^{-6}$ | $2.8 \times 10^{-6}$ | 0.998 |
| Loss Tangent | $9.5 \times 10^{-6}$ | $1.18 \times 10^{-5}$ | 0.997 |
| Total | $1.2 \times 10^{-5}$ | $1.48 \times 10^{-5}$ | $>0.99$ |

We propose a fully connected (dense) deep neural network split into two parts that outputs tunability and loss tangent separately, given frequency, electric field, $\xi_s$, $\xi_{Mg}$ and barium proportion as input features. This architecture makes it possible to use the experimental database to retrain the layers that are associated with loss tangent, without affecting the tunability. The selection criteria of number of layers and neurons in each layer is described in Appendix, section D.1.[41] Fig. 5 shows the fully connected DNN architecture which includes four hidden layers. Tunability prediction and loss tangent prediction share the input layer and the immediate dense layer. Thenceforth, the network branches off into two parts. Exponential linear unit (ELU) is chosen over rectified liner unit (ReLU) as the activation function to avoid the vanishing gradient problem and speed up the training process.[42] Linear activation is used at the output layer. As the simulated dataset was generated for discrete frequencies from 1kHz to 10GHz, we convert each of these frequencies to a one-hot vector (Appendix, section D.3 displays the one-hot vector table). This process is known as one-hot encoding where categorical data is converted into a group of bits with a single high (1) bit and all others low (0).

The DNN was implemented in python using keras-2.2.4 library with tensorflow-1.13 backend and the training was done on a RTX 2080 Ti GPU with 11GB memory. $L_2$ regularization was introduced in all layers except the last two layers of the network to prevent overfitting. K-fold cross validation (K=2) was performed to evaluate the models by having 3 separate databases for training, validation and testing (see Appendix, section D.2). Total validation loss settled around $1.48 \times 10^{-5}$ after about 6 hours of training. Table D.1 shows training and validation mean squared errors (MSEs) and coefficient of determination ($R^2$) of tunability and loss tangent predictions separately. $R^2$ value is a statistical measure that represents the goodness of a fit of a regression model. In order to validate the DL predictions, we predict the tunability and loss tangent of BST compositions that are not present in the database and compare them with the simulations. A composition depends on $x$, $\xi_s$ and $\xi_{Mg}$ and thus, out-of-database predictions consists of compositions with different values for above quantities that are unseen by the DL model. Fig. 6 demonstrates theoretical model simulation results and DL predictions for different BST-MgO composites at different frequencies. The accuracy of this DL regression model can be numerically expressed with high $R^2$ values of both tunability and loss tangent predictions and it becomes quite evident that deep learning can perfectly emulate the theoretical model.

**Table 2** Calculated $\xi_s$ and $\xi_{Mg}$ parameter values for different BST composites. All measurements and simulations are done at 20kV/cm biasing field. $\xi_{Mg}$ increases with MgO concentration for a particular composite.

| Material | Frequency | Tunability (Measurement) | Tunability (Theoretical) | Calculated $\xi_s$ | Calculated $\xi_{Mg}$ |
|---|---|---|---|---|---|
| Ba0.7Sr0.3TiO3+2.5mol%MgO [43] | 10kHz | 0.34 | 0.342 | 0.2 | 0.48 |
| Ba0.7Sr0.3TiO3+7.5mol%MgO [43] | 10kHz | 0.26 | 0.26 | 0.2 | 0.69 |
| Ba0.7Sr0.3TiO3+10mol%MgO [43] | 10kHz | 0.22 | 0.224 | 0.2 | 0.8 |
| Ba0.6Sr0.4TiO3+10wt%MgO [35] | 1MHz | 0.166 | 0.165 | 0.28 | 0.74 |
| Ba0.6Sr0.4TiO3+30wt%MgO [35] | 1MHz | 0.148 | 0.146 | 0.28 | 0.81 |
| Ba0.6Sr0.4TiO3+60wt% MgO [35] | 1MHz | 0.099 | 0.097 | 0.28 | 1 |
| Ba0.45Sr0.55TiO3 [35] | 10GHz | 0.152 | 0.145 | 0.13 | 0 |
| Ba0.5Sr0.5TiO3 [35] | 10GHz | 0.25 | 0.25 | 0.2 | 0 |

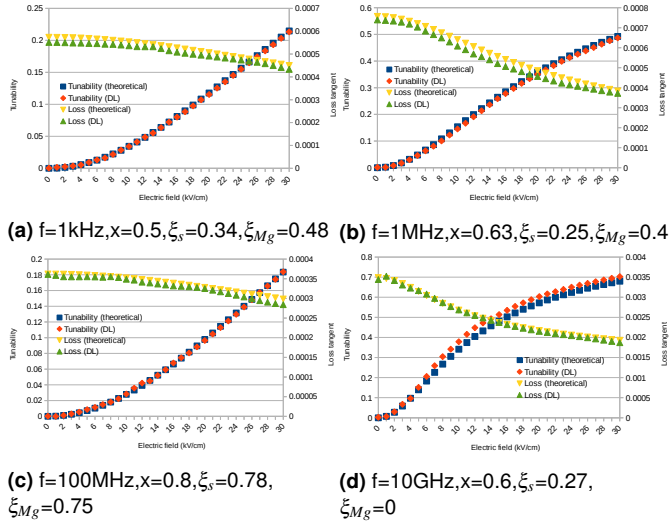

(a) f=1kHz, x=0.5, $\xi_s$=0.34, $\xi_{Mg}$=0.48
(b) f=1MHz, x=0.63, $\xi_s$=0.25, $\xi_{Mg}$=0.4
(c) f=100MHz, x=0.8, $\xi_s$=0.78, $\xi_{Mg}$=0.75
(d) f=10GHz, x=0.6, $\xi_s$=0.27, $\xi_{Mg}$=0

**Fig. 6** Theoretical model simulations and deep learning predictions for different BST+MgO composites at different frequencies. Fig. 6d represents pure BST64 material.

**Transfer learning with the measurement data**

In the second phase, we enable the pre-trained DL model to learn from the measurement dataset. In order for the database to be compatible for training, we obtained the equivalent $\xi_s$ and $\xi_{Mg}$ parameters for each of the BST composite present in the dataset by comparing measured tunability values with the theoretical model calculations. Table 2 shows the calculated $\xi_s$ and $\xi_{Mg}$ values for selected BST composites. This completed database is then utilized to learn and improve the loss tangent prediction.

The phenomenon termed "catastrophic forgetting" specifically happens when a pre-trained neural network is trained with another dataset using gradient descent algorithm as the new weight updates may not reflect previously learned features. In order to address this issue, we freeze all layers except last two layers of loss tangent prediction as shown in Fig. 5. Once a layer is frozen, it becomes non-trainable and the weights do not update upon training. Therefore, the weights of the layers associated with tunability do not update and hence the neural network will completely remember tunability characteristics learned from the theoretical model. The first three layers associated with loss tangent will remember the behaviour of loss tangent and will enable learning experimental features by training the last two layers on the measurement data.

We first filter out pure BST and only MgO-doped BST composites from the experimental database. The resulting database contains 170 data on 38 materials out of which 11 materials were selected for the test set. Due to the limitation of data, we retrain two unfrozen layers for low number of epochs, following the early stopping method, in order to prevent overfitting. An overfitted neural network performs well on the training set but has a very poor generalization accuracy. Fig. 7 shows the loss tangent prediction results on the test set. It can be observed that in all the cases, DL predictions are closer to the measurement data rather than the simulated values. For quantification purposes, we introduce a similarity score $s(p,q)$, between two sets $p$, $q$. We calculate the mean Euclidean distance $d(p,q)$ between set $p$ and set $q$ each having $n$ elements with shape $m$ as: $d(p,q) = \frac{\sum_{i=1}^{n}\sqrt{\sum_{j=1}^{m}(p_{ij}-q_{ij})^2}}{n}$ (8). Thus, similarity score, $s(p,q)$ can be introduced as the inverse of mean Euclidean distance: $s(p,q) = \frac{1}{d(p,q)}$ (9).

Calculated similarity score between the theoretical simulations and measurements is 142.2, whereas that between the DL predictions and measurements is as high as 675.6. Hence, we can conclude that deep learning offers about 4 times performance improvement in predicting the loss tangent. While we understand the measurement values can differ significantly depending on the synthesis conditions, it is still essential to develop a model that fits well with the existing data and the proposed DL model shows a better agreement with the experimental measurements.

**Interactive learning framework**

New predictions from the 'transfer-learned' DL model assist the human to interactively make appropriate adjustments to the theoretical model parameters. As shown in Fig. 8, the interactive learning work flow is a reciprocal process done in two cycles. In step 1, theoretical simulations (tunability and loss tangent) are carried out to be compared with the experimental data. Then in step 2, a comparison is done by manual inspection and the theoretical model parameters are tweaked heuristically. In the real scenario, by manual comparison with the results from Ref. [44,45], we found that the calculated loss tangent at low frequencies (1kHz - 100kHz) are generally much lower estimated. In the original model, the resonance frequency of the low frequency relaxation loss $\Gamma_4$ is set to 10MHz. Therefore, we propose a new low frequency relaxation formula of $\Gamma_5$ resonant at $f_5 =$ 10kHz and $\omega_5 = 2\pi f_5$ as: $\Gamma_5 = A_5/(1 - i\omega/\omega_5)$ (10), which corrects the underestimation given by the model. The parameter $A_5$ is assumed to be equal to the low frequency relaxation parameter of the original model (see Appendix, section A.3).

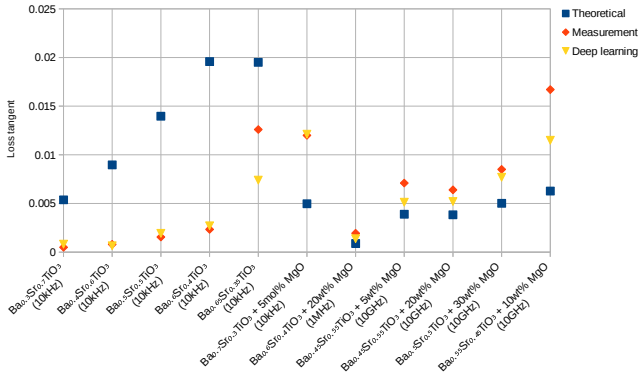

**Fig. 7** Comparison of theoretical model simulations, measurements and deep learning predictions on the test set (E=20kV/cm). It should be noted that the theoretical simulations are carried out after the human-machine interactive learning improvements.

However, the first learning cycle limits ourselves only to the existing data. Therefore, we make use of experimental-data-trained DL model to predict the tunability and loss tangent of new ferroelectric-dielectric composites that are to be compared with the simulations. Theoretical model generated data in step 3 are compared with the DL predictions in step 4 and the model parameters are again tuned to confront with the predictions. In the actual case, we observed that DL predicted loss tangent at 1MHz is over 10 times larger than that of the simulated value. This was confirmed by doing further literature review and finding the corresponding experimental value.[35] Hence, the subsequent fine-tuning is performed on the theoretical model parameters. In our previous simulations, all model constants were set to be the same as original Vendik's model in Ref.[8], referring to Appendix, section A.3. Since the loss tangent was obviously lower estimated around 1 MHz, we increased the value of coefficient of low-frequency loss $A_4$. After some heuristic tweaking, here we assign a new value 0.01 for $A_4$.

The first learning cycle could be referred to as a manual human learning procedure whereas the second could be identified as a human - machine learning interaction, since the human adjusts model parameters depending on the feedback of a trained DL model. At the end of two learning cycles, the theoretical model has adjusted to the measurement data as well as possible.

**DL optimized materials modelling**

We regard an objective function to consider both tunability ($n_r$) and loss tangent ($\tan\delta$) factors such that optimal material properties can be quantified. Here, we define the figure of merit (FOM/K)[46,47] factor as: $K = \frac{n_r}{\tan\delta}$ (11). Using the trained DL model, we investigate the best FOM values at different frequencies and the corresponding barium proportions and defect parameter values. For the sake of convenience, we perform this for pure BST materials. In the present paper, we investigate FOM at 20 kV/cm, which is the most frequent value present in our literature data. We range the proportion of barium from 0.5 to 0.7 (assuming the material is at paraelectric state in room temperature when x≤0.7[48]) and the defect factor $\xi_s$ from 0.2 to 0.8. It should be noted that for all practical BST ceramics, there is always some existing defects, and we assume that the initial lowest value of defect factor $\xi_s$ is 0.2.[49] Several example frequencies were selected from 100kHz to 10GHz for the proposed optimization and the temperature is set to be the room temperature at 290K, at which the dataset was generated. The corresponding combination of x and $\xi_s$ which results the highest FOM value can theoretically be regarded as the best BST material under the considered frequency and temperature.

However, the predicted loss tangent being too low can result in very high FOM values without revealing much information about the tunability and the overall performance of the material. Hence, while calculating the FOM value from the DL model, we set the minimum threshold of the loss tangent to be $5\times10^{-4}$, as it is the lowest loss value observed in the experimental dataset.[50] Table 3 shows the best FOM values and the corresponding $x$ and $\xi_s$ values obtained using the DL predictions at different frequencies for pure BST materials. By observing the DL results, it can be concluded that the best operating frequency for pure BST materials is around 10MHz as it shows the highest K value among the chosen frequencies. The optimum barium proportion is found to be around 0.63 for most of the frequencies. From Table 3, it can also be noticed that low level of defects are preferred in microwave frequencies whereas relatively high defects provide better FOM values in low frequencies.

**Table 3** Best figure of merit obtained from DL model at different frequencies at 20 kV/cm.

| $f$ | 100kHz | 10MHz | 1GHz | 10GHz |
|---|---|---|---|---|
| x | 0.54 | 0.63 | 0.6 | 0.63 |
| $\xi_s$ | 0.68 | 0.70 | 0.20 | 0.20 |
| FOM | 273.38 | 551.49 | 35.66 | 14.04 |

**Experimental validation**

BST64 samples sintered at different temperatures were taken as an example material to be studied (see Appendix, section C for sample preparation methods). Depending on these different synthesis conditions, the defects may differ for each sample and it is worth investigating whether our model could capture the correct defect parameter $\xi_s$, and the corresponding dielectric properties of these samples. For all the prepared BST64 samples, the dielectric constant was measured from 250K to 400K respectively at 100kHz under zero external biasing field. Therefore, we perform dielectric constant vs temperature simulations at 100kHz while keeping E=0kV/cm. Fig. 9 shows the best curve fittings between experimental data (circles) and the simulation data (lines) for BST64 synthesized with different sintering methods and temperatures. Through the fitting process, we can obtain unique values of $\xi_s$ for each type of material and the simulation results match quite well with the measurements, especially above the Curie point. We believe that the observed poor fit below the phase transition temperature is due to the simplified calculation of Landau-Ginzburg equations used to derive the Vendik model (see Appendix, section C). Moreover, microstructures of BST64 pellets at different sintering conditions were investigated by using FEI Quanta FEG 400 high resolution scanning electron microscope. Table 4 shows the grain sizes of different samples and the calculated $\xi_s$ values. High $R^2$ values evidence that our model is able to capture the correct dielectric properties of different BST64 materials having different defects. Comparing values of estimated average grain sizes and defect factors, no clear trend can be concluded as the data size is not enough. However, for samples synthesised by SPS process, grain sizes are obviously smaller and values of defect factor $\xi_s$ are higher, indicating a greater density of defects in BST material.

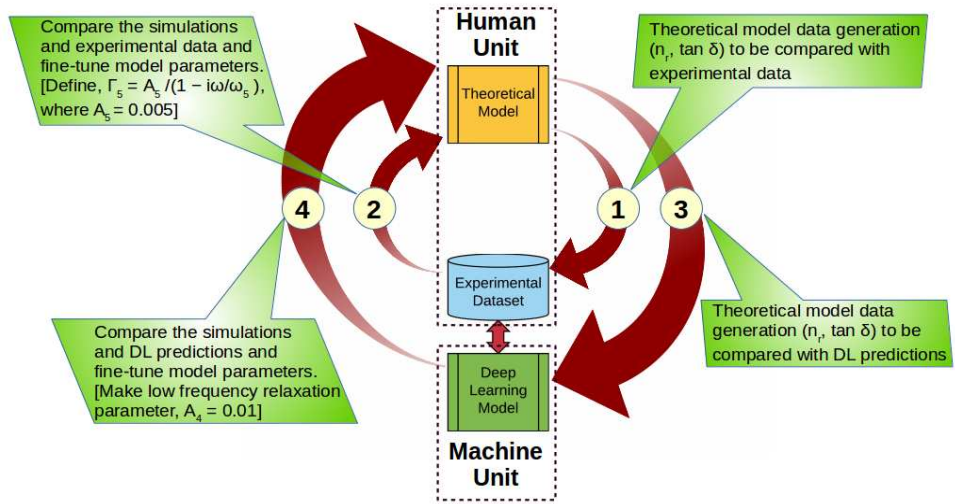

**Fig. 8** BST material modelling using human - machine learning interaction. In step 1, theoretical simulations are performed to be compared with the experimental data. In step 2, model parameters are tweaked accordingly by manual inspection and comparison. Step 3 refers to the data generation that are to be compared with machine-generated DL predictions. A human compares these two data and make appropriate refinements to the model parameters in step 4.

**Table 4** Obtained parameters from simulations for BST64 materials sintered at different temperatures, pressures and time intervals.

| Synthesis | Calculated $\xi_s$ | Grain size($\mu$m) | $R^2$ Value |
|---|---|---|---|
| SPS (1150/5m) | 0.59 | 0.5 | 0.9834 |
| CS (1200/3h) | 0.18 | 0.8 | 0.9190 |
| CS (1300/3h) | 0.20 | 1.2 | 0.9712 |
| CS (1400/3h) | 0.22 | 10 | 0.9728 |
| CS (1500/3h) | 0.19 | 25 | 0.9771 |

## Conclusions

A new framework of human-machine interactive learning has been developed for accurate modelling of ferroelectric-dielectric composites. By integrating big data generated from a semi-empirical model and the measurement database of sufficient size, we have trained a DL model, which refinement of a classical model of ferroelectric materials can be made to account for multiple unknowns. The model was experimentally validated with BST64 samples synthesised at different sintering conditions and the simulations show a good agreement with the measurements. We believe that this approach has a far reaching implication for applications in discovering new material models, especially those analytically unsolvable. As future work, we plan to apply the developed DL model to automate the materials design process as well as perform a thorough analysis on the dependence of less-studied $\xi_s$ and $\xi_{Mg}$ parameters on grain sizes, domain walls and oxygen vacancies in ferroelectric-dielectric composites.

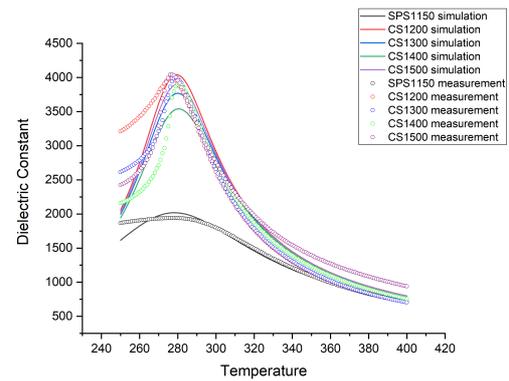

**Fig. 9** Best fitting of dielectric constant versus temperature between simulation data (line) and experimental data for pure BST64 materials (at 100kHz) sintered at different conditions. SPS - spark plasma sintering; CS - conventional sintering.

## Conflicts of interest

There are no conflicts to declare.

## Acknowledgements

We thank Ms. Hanchi Ruan for collecting data. This work is supported in part by "Software Defined Materials for Dynamic Control of Electromagnetic Waves" (ANIMATE) project (Grant No. EP/R035393/1) and the authors acknowledge Engineering and Physical Sciences Research Council (EPSRC) for providing funding for AOTOMAT (Grant No. EP/P005578/1), TERRA (Grant No. EP/S010009/1), TERALINKS (Grant No. EP/P016421/1) and SYMETA (Grant No. EP/N010493/1). A.I. acknowledges IET AF Harvey Research Prize for funding the PhD studentship.

## Notes and references